\documentclass[10pt,aps,prc,twocolumn,floatfix,amsmath,amssymb,nofootinbib,superscriptaddress]{revtex4-1}
\usepackage{color}
\usepackage{graphicx}
\usepackage{dcolumn}    
\usepackage{multirow}
\usepackage{bm}         
\usepackage{sidecap}
\usepackage{hyperref}   
\usepackage{amsmath}
\usepackage{graphicx}
\usepackage{adjustbox}
\usepackage{hyperref}                

\usepackage{mathtools} 

\usepackage[dvipsnames,usenames]{xcolor}
\usepackage{mathrsfs,natbib}
\usepackage{epsf,amssymb,amsbsy,amsfonts,amssymb,amsmath}
\usepackage{slashed}
\usepackage{comment}

\newcommand{\fmiq}{\, \text{fm}^{-3}}

\newcommand{\mm}{\mathrm{MM}}
\newcommand{\nr}{\mathrm{NR}}
\newcommand{\rl}{\mathrm{RL}}
\newcommand{\sat}{\mathrm{sat}}
\newcommand{\sym}{\mathrm{sym}}
\newcommand{\nuc}{\mathrm{nuc}}
\newcommand{\Qyc}{\mathrm{Qyc}}

\newcommand{\QCD}{\mathrm{QCD}}

\usepackage{hyperref}
\usepackage[justification=raggedright]{caption}
\definecolor{dark-red}{rgb}{0.,0.,0}
\definecolor{dark-blue}{rgb}{0.,0.,1}
\definecolor{medium-blue}{rgb}{0,0,1}
\hypersetup{
  colorlinks, linkcolor={dark-red},
  citecolor={dark-blue}, urlcolor={medium-blue}
}

\begin{document}

\title{Quarkyonic stars with isospin-flavor asymmetry}

\author{J\'er\^ome Margueron} \affiliation{Univ Lyon, Univ Claude
  Bernard Lyon 1, CNRS, IP2I Lyon / IN2P3, UMR 5822, F-69622,
  Villeurbanne, France}

\author{Hubert Hansen} \affiliation{Univ Lyon, Univ Claude Bernard
  Lyon 1, CNRS, IP2I Lyon / IN2P3, UMR 5822, F-69622, Villeurbanne,
  France}

\author{Paul Proust} \affiliation{Univ Lyon, Univ Claude Bernard Lyon
  1, CNRS, IP2I Lyon / IN2P3, UMR 5822, F-69622, Villeurbanne, France}

\author{Guy Chanfray} \affiliation{Univ Lyon, Univ Claude Bernard Lyon
  1, CNRS, IP2I Lyon / IN2P3, UMR 5822, F-69622, Villeurbanne, France}

\date{\today}

\begin{abstract}
  We suggest an extension to isospin asymmetric matter of the quarkyonic model 
  from McLerran and Reddy~\cite{McLerran:2019}. 
  This extension allows us to construct the $\beta$-equilibrium between quarks, 
  nucleons and leptons.
  The concept of the quarkyonic matter originates from the large number of color limit 
  for which nucleons are the correct degrees of freedom near the Fermi surface -- reflecting 
  the confining forces -- while deep inside the Fermi sea quarks naturally appear.
  In isospin asymmetric matter, we suggest that this new concept can be implemented
  within a global isoscalar relation between the shell gaps differentiating the nucleon 
  and the quark sectors.
  In addition, we impose the conservation of the isospin-flavor asymmetry in the 
  nucleon and the quark phases.
  Within this model, several quarkyonic stars are constructed on top of the
  SLy4 model for the nucleon sector, producing a bump
  in the sound speed, which implies that quarkyonic stars are systematically bigger 
  and have a larger maximum mass than the associated neutron stars.
  They also predict lower proton fraction at $\beta$-equilibrium, which potentially quenches 
  fast cooling in massive compact stars.
\end{abstract}

\maketitle

Recent observations of neutron stars (NS), such as radio and x-ray
astronomy, or the detection of gravitational wave (GW) detection, have
provided the tightest constraints on the dense matter equation of
state (EoS) to date~\cite{NewCompStar2019,HPY:2007,Ligo:2017}.
These constraints can be classified into different groups:
the first one refers to the highest NS masses ever
observed~\cite{Antoniadis:2013,Ozel:2016,Fonseca:2016,Linares:2018,Cromartie:2018,Arzoumanian:2018},
estimated to be about $2M_\odot$, with some indications that the
maximum mass could eventually be
larger~\cite{Cromartie:2018,Arzoumanian:2018}.
The second group assembles constraints from binary NS
(BNS) GW detection in the in-spiral phase, from which the tidal deformability is
estimated~\cite{Hinderer:2008,Flanagan:2008} .
It includes GW170817~\cite{Ligo:2017} and further detections.
The third group still refers to NS mergers, more specifically to the
analysis of the electromagnetic (EM) counterpart, see
Ref.~\cite{Coughlin:2019} for instance and refs. therein.
The fourth group matters with x-ray observations, such as thermal
emission from qLMXB~\cite{Guillot:2013,Steiner:2018,Baillot:2019},
x-ray burst and photospheric expansion~\cite{Ozel:2016,Carolyn:2016},
as well as x-ray emission from hot-spots at the surface of some NS
(NICER)~\cite{NICER}.

The analysis of these new data requires to set-up generic models for the
EoS, with various levels of agnosticity to model assumptions, such as
the isospin asymmetry for instance, or the interaction prerequisites, inducing
possible spurious constraints among observables.
As an illustration, the nuclear meta-model~\cite{Margueron:2018a} can be used
to explore predictions compatible with the assumption
that matter is uniquely composed of nucleons (and of course leptons).
It can also be employed for the description of phase transitions. 
This is the purpose of this paper: exploring the nucleon-quark phase 
transition in compact stars where this transition is described 
with the quarkyonic model~\cite{McLerran:2019}.

The quarkyonic model for dense matter proposed in Ref.~\cite{McLerran:2007} 
(and recently applied to neutron stars in Ref.~\cite{McLerran:2019})
is one of them. It is an interesting candidate to bridge the gap in
describing quark matter and nuclear matter at the phase transition
\cite{Fukushima:2016}.
The quarkyonic model is not properly a microscopic model since it is not based on the
QCD Lagrangian or an effective version of it, but it implements some
features from in the large number of color ($N_c$) limit of
QCD~\cite{McLerran:2007} with few parameters.
New configurations at high and low temperature limit of the
holographic Witten-Sakai-Sugimoto model has recently been interpreted
as holographic realizations of quarkyonic matter in isospin symmetric matter, 
based on a quark Fermi sea enclosed by a baryonic layer on momentum
space~\cite{Kovensky2020}. In the real-world where $N_c=3$, it
possibly approximates the actual ground state of dense matter and the
confrontation of its predictions to the data can be used to determine
the model parameters.
These parameters happen to be physical and thus meaningful: they
are the quarkyonic scale $\Lambda_{\Qyc}\approx 250-300$~MeV, which is
comparable to the QCD scale $\Lambda_{\QCD}$, and $\kappa_{\Qyc}$
which controls the saturation of the nucleonic shell~\cite{McLerran:2007}.

The interesting feature of the quarkyonic model is that it suggests a
cross-over between the hadron phase and the quark one, at
variance with other approaches such as the ones based on
Maxwell/Gibbs construction.
Note that cross-over are also suggested from Cooper pairing in the hadron and in 
the quark phases. The quarkyonic cross-over is thus another example
of such features.
It however disregards one of the essential prediction of QCD, namely the
restoration -- as the density increases -- of chiral symmetry which is 
spontaneously broken in the QCD vacuum.
Note that in the holographic approach from Ref.~\cite{Kovensky2020}, chirally 
restored and chirally broken quarkyonic matter are constructed, and it was found
that only chirally restored matter is energetically preferred.
In widely used quark models implementing the chiral symmetry breaking, e.g.
Nambu--Jona-Lasinio approach, the transition between the broken phase assimilated
to the hadronic phase and the restored one assimilated to the free quark phase 
is however generally first order (for most parameter sets).
In the future, it would be interesting to combine together in isospin asymmetric matter
the phenomenology of the color gauge symmetry realized at large $N_c$ and the 
chiral symmetry dynamics, both rooting into the QCD Lagrangian.

In the cross-over region, the quarkyonic model suggests that the
pressure first increases at the onset of the first quarks, while first
order phase transition models usually suggest a softening of the
pressure due to the increase of the degrees of
freedom~\cite{Zdunik:2013,Alford:2013}. The consequence is the large
increase of the energy density, as well as a peaked sound speed at a
density of about 2-3 times the saturation density of nuclear matter,
$n_\sat\approx 0.16\fmiq$, as expected by some
authors~\cite{Steiner:2015,Tews:2018}.

These features are characteristic to the quarkyonic model -- however not specific --
and have motivated further investigations and applications to neutron star physics. 
In the original paper by
McLerran and Reddy~\cite{McLerran:2019}, the quarkyonic matter was
studied in the case of isospin symmetry (symmetric matter, SM) as well as
in the case of neutron matter (NM). The application to neutron star has
been performed assuming it is composed only of neutrons, and $u$ and
$d$ quarks within the ratio satisfying local charge neutrality,
$k_{Fd}=2^{1/3}k_{Fu}$ where $k_{Fd}$ ($k_{Fu}$) is the $d$ ($u$)
quark Fermi momentum. A version of the quarkyonic model for isospin asymmetric
matter, where the isospin asymmetry is controlled by the chemical
equilibrium, was then suggested by Zhao and Lattimer~\cite{Zhao:2020}.
In their model, Zhao and Lattimer have treated nucleons and quarks as
independent particles for which the energy minimization imposes the
equilibrium between their respective chemical potential. They found
that their quarkyonic stellar model is able to satisfy observed
mass and radius constraints with a wide range of model parameters.
They also predict that quarkyonic matter tends to reduce the proton
fraction, compared to the nucleonic case. This reduces the domain of
parameter allowing the direct URCA process~\cite{Lattimer:1991}.
It was also suggested by Jeong, McLerran and Sen~\cite{Jeong:2020}
that the hard core in the nucleon interaction could be represented by
an excluded volume, which in turns can be related to the shell gap
controlling the cross-over properties. In such model, the shell gap is
directly controlled by the size of the hard core. Similarly to the
original paper by McLerran and Reddy~\cite{McLerran:2019}, this model
also predicts the presence of a peak in the sound speed at
2-3$n_\sat$.
It was also extended to describe three-flavor baryon-quark mixtures, allowing
the onset of strange particles~\cite{Duarte:2020a,Duarte:2020b}.

In this paper, we suggest another version of the quarkyonic model for isospin
asymmetric matter (AM) where we investigate the analogy between the quarkyonic model 
and the Cooper pair formation around the Fermi energy~\cite{McLerran:2019}. 
While this analogy may appear as rather simplistic, it suggests that quarks 
and nucleons may be viewed as two representations of the same quasi-particle
excitation, in the same way as Cooper pairs and single particles coexist in superfluids 
or superconductors, but in different regions of the nuclear spectrum~\cite{deGennes}.
In AM, the neutron/proton ratio in the nucleon sector and the flavor 
asymmetry in the quark sector are fixed by the compound nature
of the nucleons, since $n: udd$ and $p: uud$.
In this spirit, quarks and nucleons are not distinguished as two independent
particles for which the energy minimization imposes an equilibrium relation, 
as suggested in  the quarkyonic model of Zhao and Lattimer~\cite{Zhao:2020}.
In addition, the $\beta$-equilibrium does not involve quark chemical potentials
since only nucleons are occupying the Fermi levels.
Our picture requires a new approach for the thermodynamical construction 
of the phase equilibrium. 
The cross-over, as described in the original quarkyonic 
model~\cite{McLerran:2019}, is depicted by an isoscalar condition 
connecting the momenta of the quarks and of the nucleons, while the isospin/flavor
asymmetry in the quark and the nucleon sectors is fixed.
Under these two assumptions, the model we propose describes the phase transition 
from symmetric to neutron matter.

In our picture, there is no direct contribution of the quarks to the $\beta$-equilibrium 
since they do not occupy Fermi levels. The presence of quarks however influences
the $\beta$-equilibrium, through their contribution to the nucleon chemical potentials.
This picture breaks down in the pure quark phase, which does not occurs in the 
quarkyonic model since there is always a small but finite contribution of nucleons at
high density.
In addition the chiral symmetry generating the constituent quark mass is assumed 
to remain at all density, even in the dense phases where quarks become the 
dominant species. 
This is also an interesting suggestion from the quarkyonic model which goes 
against the usual picture of the hadron-quark phase transition based on chiral 
symmetry restoration. 
As we stated earlier, on the one hand the quarkyonic transition is driven
by features of QCD relying on its gauge theory nature at large $N_c$,
where only planar graphs survive, whereas on the other hand the
transition follows the chiral symmetry restoration (property of the
quark sector only) which induces a large change of the constituent
masses and also of the baryon properties. 
In the future, a model unifying both mechanisms in isospin asymmetric matter
would be an interesting theoretical development.

\medskip

In this paper, we suggest an extension of the original quarkyonic
model~\cite{McLerran:2019} for AM in Sec.~\ref{sec:qam}.
The cold catalyzed NS EoS -- at $\beta$-equilibrium -- is derived
in Sec.~\ref{sec:qbm} and finally we calculate 
NS properties in Sec.~\ref{sec:ns}.
We finally conclude and suggest outlook in Sec.~\ref{sec:conclusions}.

\section{Quarkyonic model in asymmetric matter}
\label{sec:qam}

The concept of quarkyonic matter has emerged in the large number of
color, $N_c$, limit of QCD~\cite{McLerran:2007}.
In this limit and when the nucleon density $n_N$ is much larger than
the QCD scale, $n_N\gg\Lambda_\QCD^3$, the confining potential of QCD
is dominant even though 
the nucleonic Fermi momentum is large, $k_{F_N}\gg\Lambda_\QCD$. 
The concept of quarkyonic matter has been introduced in order to
resolve this apparent paradox:
the ground state of dense matter is composed
of dressed quarks (with mass $\simeq M_N/3$) that are freely moving
deep inside the Fermi sea, and of a shell of baryons generated by the
strong confining force, which lies
close to the Fermi level~\cite{McLerran:2007}.
Baryons occupies a momentum shell whose width is taken to be 
$\Delta_\Qyc\approx \Lambda_{\Qyc}$, while deepest states are 
occupied by quarks.

\begin{figure}[t]
  \begin{center}
    \includegraphics[width=0.45\columnwidth]{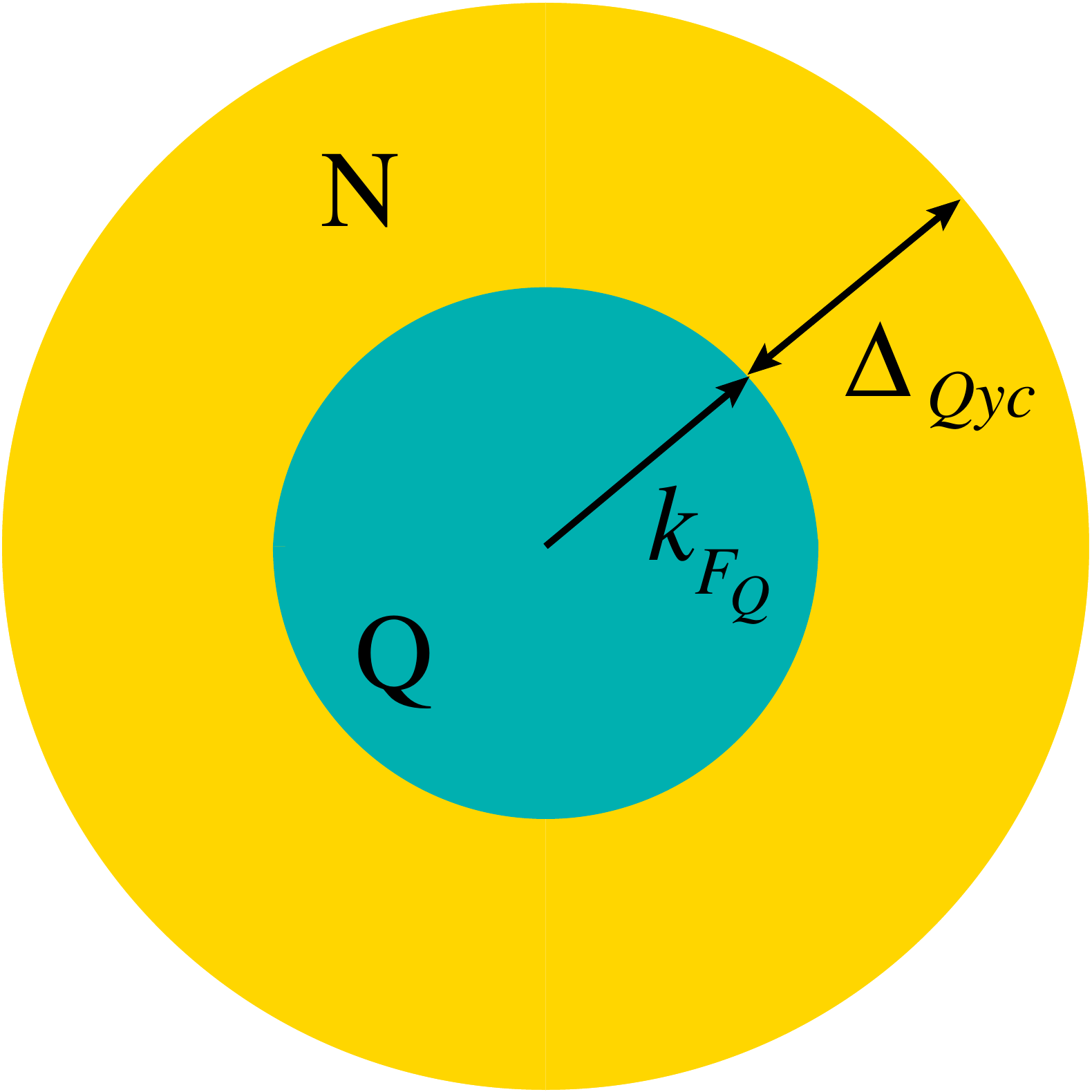}
    \includegraphics[width=0.45\columnwidth]{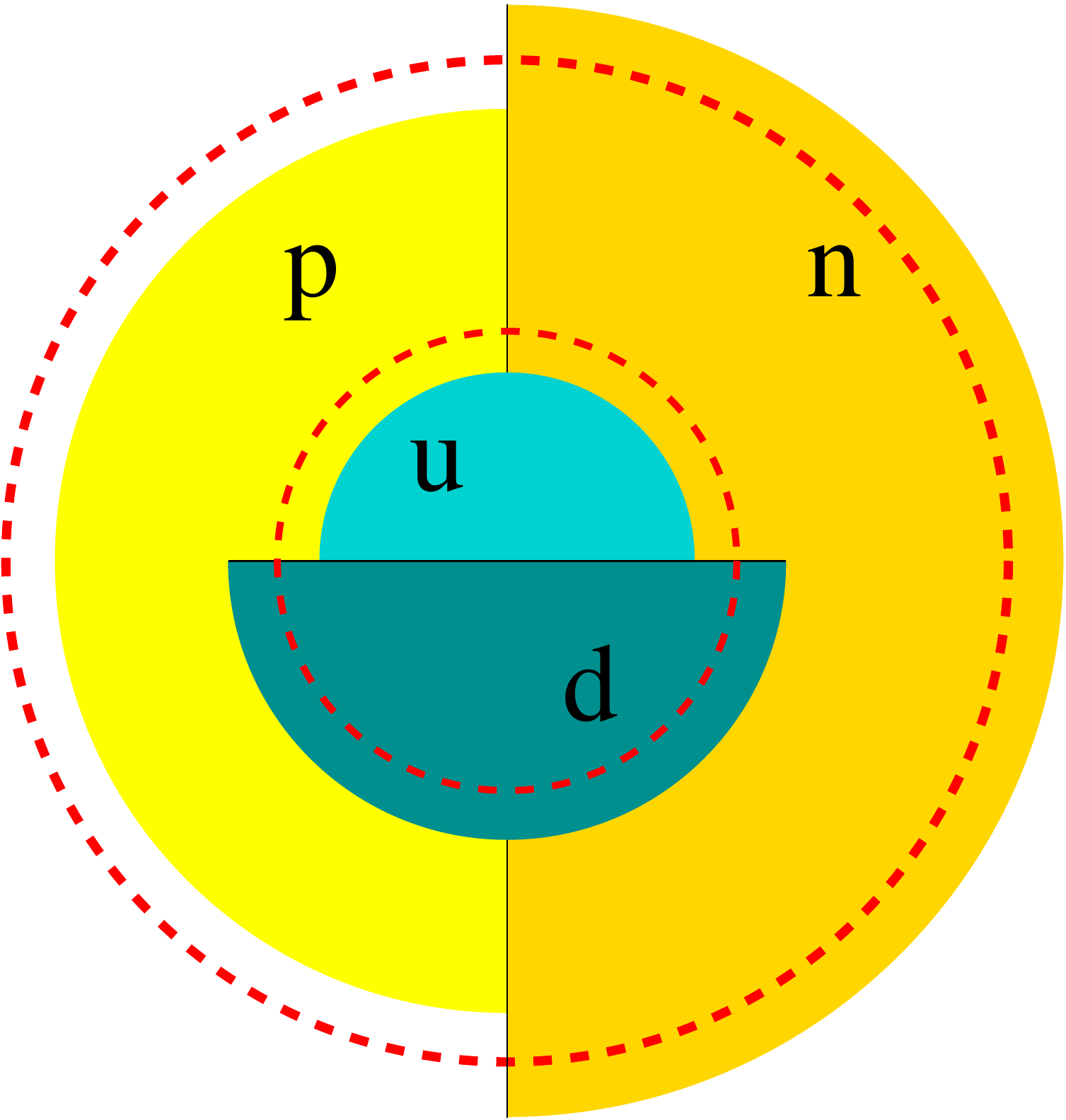}
    \caption{Schematic views of the different Fermi seas at stake in quarkyonic 
    matter for SM (left) and AM (right). Quarks occupy deep states
      inside their Fermi spheres while nucleons occupy the shells close 
      to their Fermi levels. }
    \label{fig:FermiSeas}
  \end{center}
\end{figure}

Even if matter is composed of nucleons and quarks, low energy
excitations around the Fermi level involve only quasi-particles of
nucleonic type. The excitation of quarks require the expense of a
momentum of the order of the shell width $\Delta_\Qyc$. 
In this prospect, nucleons and quarks are not independent particles,
but there are realization of matter excitations in well separated
energy regions.
This picture is clearly illustrated on the left panel of Fig.~\ref{fig:FermiSeas} 
for SM.

In SM, this picture is simple since it is sufficient to
fix a condition between the nucleon and quark single particle energies
$\epsilon_N(k)$ and $\epsilon_Q(k)$ to separate the deep quark
states from the nucleons ones located around the Fermi level. This
condition imposes that the last quark occupied state coincides with
the first nucleon one~\cite{McLerran:2019},
\begin{equation}
\epsilon_N^\mathrm{NI}(k_{F_N}-\Delta_\Qyc) = N_c \epsilon_Q(k_{F_Q})
\label{eq:spenucq}
\end{equation}
where $k_{F_N}$ and $k_{F_Q}$ are isoscalar nucleon and quark Fermi momenta.
The factor $N_c$ in Eq.~\eqref{eq:spenucq} stands
for the number of quarks forming each nucleon in its ground
state. With this condition, the nucleon and quark degrees of freedom
are reduced to only one free variable, that we fix to be $k_{F_N}$, the 
nucleon Fermi momentum for convenience.
Note that in Eq.~\eqref{eq:spenucq} the nucleon single particle energy is
the non-interacting (NI) one,
$\mathcal{E}_N^\mathrm{NI}(k)=\sqrt{M_N^2+k^2}$.  This choice is
performed in order to avoid unnecessary complication of the model.
We further assume that chiral symmetry remains broken to set the
quark mass ($M_u=M_d=M_Q$) with $M_Q \approx M_N/N_c$, as in the
constituent quark model.

Finally a prescription for the thickness of Fermi layer where nucleons
reside has to be taken. We adopt the same relation from the original 
paper~\cite{McLerran:2019},
\begin{eqnarray}
  \Delta_{\Qyc}= \frac{\Lambda_{\Qyc}^3}{\hbar c^3k_{F_N}^2} + \kappa_{\Qyc} \frac{\Lambda_{\Qyc}}{\hbar c N_c^2} \, ,
  \label{eq:delta}
\end{eqnarray}
with $\Lambda_{\Qyc}\approx 250-300$~MeV and
$\kappa_{\Qyc}\approx 0.3$. 
$\Delta_{\Qyc}$ defines the energy scale below which nucleons reside.

The concept of quarkyonic matter however leads to a fundamentally new 
way to represent the nucleon and quark densities $n_N$ and $n_Q$ 
and their associated Fermi seas.
As in a superfluid, there is only one chemical potential that enters
into the thermodynamic equilibrium, and which is associated to the
last nucleon occupied state $\mu_N=E(N_B)-E(N_B-1)$ where $E$ is the
total baryon energy in the ground state and $N_B$ stands for the 
number of baryons which is a partition of nucleons and quarks states.
By adding or removing a baryon, the entire Fermi sea is reorganized
leading to a new partition between quark and nucleon states, with the
condition on the number densities $n_B=n_N+n_Q$. This new concept is
first applied to symmetric and neutron matter~\cite{McLerran:2019} and
we are now suggesting an extension for AM.

\subsection{Global isoscalar relation between nucleon and quark Fermi seas}

By breaking the isospin symmetry in AM, the nucleon and quark states 
are replaced by four other states: neutrons, protons, as well as $u$ and $d$ quarks
are the natural components in AM, where they are represented
by their associated densities, $n_n$, $n_p$, $n_u$ and $n_d$.
The four Fermi seas are schematically represented in the right panel of
Fig.~\ref{fig:FermiSeas}. 
The baryon charge density controlling SM is completed in AM by the
isospin asymmetry $\delta_N$.
Since there are only two charges and four particles, the concept of quarkyonic 
matter in asymmetric matter requires two additional relations.

At variance with SM where nucleons and quark Fermi
seas could be defined by imposing a relation between their
single particle energies, see Eq.~\eqref{eq:spenucq}, nothing similar can be done in 
AM between the four particles. 
The concept of quarkyonic matter suggests however that the relation
between nucleons and quarks in SM, see Eq.~\eqref{eq:spenucq}, remains 
globally valid in AM. 
The nucleon and quark Fermi momenta in AM, $k_{F_N}$ and $k_{F_Q}$, 
are represented in Fig.~\ref{fig:FermiSeas} by the red dashed circles.

Expanding the single particle energies in Eq.~\eqref{eq:spenucq},
e.g. $\mathcal{E}_Q(k)=\sqrt{M_Q^2+k^2}$, one gets the following relation 
between the nucleon and quark isoscalar Fermi momenta,
\begin{eqnarray}
  k_{F_Q} &=& \frac{k_{F_N}-\Delta_{\Qyc}}{N_c} \Theta(k_{F_N}-\Delta_{\Qyc})  \, .
              \label{eq:kfq}
\end{eqnarray}

Note that in Eq.~(\ref{eq:kfq}) we assumed that the nucleon
shell gap $\Delta_{\Qyc}$ is also an isoscalar quantity.
Eq.~(\ref{eq:kfq}) allows quark to appear as soon as
$k_{F_N}-\Delta_{\Qyc}>0$ independently of the isospin asymmetry. This
is supported, as we explain previously, by the idea that exciting
quarks in AM requires an energy of the order of $\Delta_{\Qyc}$
irrespectively of the partition of matter between neutrons and
protons. Without an actual solution of QCD, this assumption is the
simplest one which can be done.

The limit of NM was also explored in the original
paper by McLerran and Reddy~\cite{McLerran:2019}. The prescription
taken there imposes that one has to choose which among $u$ and $d$
quarks are connected to the neutron states. 
In addition, the value for $\Lambda_\Qyc=300$~MeV considered in SM
was changed to $380$~MeV in NM.
In our model, $\Lambda_\Qyc$ is taken constant, but the change of the Fermi 
momentum between SM and NM, see Eqs.~(\ref{eq:kfq}) and (\ref{eq:kfnp}), induces an effective modification of 
$\Lambda_\Qyc$ between SM and NM with the ratio $2^{1/3}$, which is
exactly the same ratio considered by McLerran and Reddy~\cite{McLerran:2019}.
In practice, the two approaches lead to similar results in NM.
The isoscalar relation~(\ref{eq:kfq}) presents however the advantage to 
describe AM and to recover the limit of SM, where the concept of the quarkyonic 
matter is simple.

In summary, we remark that the isoscalar Fermi momentum $k_{F_N}$ controls 
both the isoscalar quark Fermi momentum $k_{F_Q}$, see Eq.~\eqref{eq:kfq}, 
as well as the nucleon gap $\eqref{eq:kfq}$ from the prescription \eqref{eq:delta}.
The isoscalar nucleon and quark densities can be safely determined as,
\begin{eqnarray}
  n_N = \frac{2}{3\pi^2}
  \Big[ k_{F_N}^3 - (k_{F_N}-\Delta_{\Qyc})^3\Theta(k_{F_N}-\Delta_{\Qyc})  \Big]  \, , \nonumber \\
  \label{eq:nnuc}
\end{eqnarray}
and the quark density,
\begin{eqnarray}
  n_Q &=& \frac{2}{3\pi^2}  k_{F_Q}^3 \Theta(k_{F_Q}) \, . \label{eq:nq}
\end{eqnarray}
The total baryon density is built upon the nucleon and quark contributions as,
\begin{eqnarray}
  n_B = n_N + n_Q \, .
\end{eqnarray}

While the isoscalar densities $n_N$ and $n_Q$ can now be calculated from $k_{F_N}$,
the connection to the densities of the four particles $n$, $p$, $u$ and $d$ are yet
unknown. They are related to the isoscalar densities from the following relation:
\begin{eqnarray}
n_N &=& n_n+n_p \, \\
n_Q &=& (n_d+n_u)/N_c \, .
\end{eqnarray}

In the following, we suggest that the densities $n_n$, $n_p$, $n_u$ and $n_d$
can be obtained in AM by imposing the isospin/flavor asymmetry in the
nucleon and quark phases.

\subsection{Isospin/flavor asymmetry}

We will now determine the particle densities.
In the quarkyonic model, there is a partition between nucleons and
quarks which changes as the density evolves.
It reveals the dynamical process converting quarks into nucleons or
breaking the nucleons into their constituents, as it shall be in 
compound systems.
A simple way to translate this symmetry into the nucleon and quark phases
is to impose that the two phases conserve the isospin/flavor asymmetry.
Since $n:(udd)$ and $p:(uud)$, we obtain the following relations for the
quark density in the nucleon phase, 
$n_u^\nuc=n_n+2n_p$ and $n_d^\nuc=2n_n+n_p$,
which leads to the following simple connection between the isospin asymmetry parameter
$\delta_N=(n_n-n_p)/n_N$ in the nucleon phase and the flavor asymmetry parameter 
$\delta_Q=(n_d-n_u)/(n_d+n_u)$ in the quark phase:
\begin{equation}
  \delta_N = N_c \delta_Q \, .
  \label{eq:deltaNQ}
\end{equation}
From SM to NM, $\delta_N$ goes from $0$ to $1$, while $\delta_Q$ goes
from $0$ to $1/N_c$.
The dynamics of the phase transition thus imposes that the $u$ and $d$
flavor ratio reflects the isospin asymmetry of the nucleon phase.

As a side note, we are aware that the isospin/flavor asymmetry relation~\eqref{eq:deltaNQ} 
can possibly be violated by the two phases if the energy minimization is injected.
Such a refinement of the quarkyonic model is indeed very interesting but it complexifies
the quarkyonic approach, which nice feature remains in its simplicity.
Further extensions of the present model will be explored in the future, especially to analyze their role in 
the predictions presented here.

Knowing $\delta_N$ and $k_{F_N}$ -- which fixes $n_N$ and $n_Q$ -- one can deduce all particle densities as
\begin{eqnarray}
  n_n &=& \frac{1+\delta_N}{2} n_N \equiv x_n n_N\, , \\
  n_p &=& \frac{1-\delta_N}{2} n_N \equiv x_p n_N \label{eq:np}\, , \\
  n_d &=& \frac{1+\delta_Q}{2} N_c n_Q \equiv x_d N_c n_Q\,  , \\
  n_u &=& \frac{1-\delta_Q}{2} N_c n_Q \equiv x_u N_c n_Q \, .
\end{eqnarray}

The $u$ and $d$ quark Fermi momenta are simply related to their
densities as
\begin{eqnarray}
  k_{F_u}^3 &=& \frac{3\pi^2}{N_c} n_u  = (1-\delta_Q) k_{F_Q}^3\, , \\
  k_{F_d}^3 &=& \frac{3\pi^2}{N_c} n_d  = (1+\delta_Q) k_{F_Q}^3\, ,
\end{eqnarray}
since $d$ and $u$ quarks occupy their Fermi sphere, see
Fig.~\ref{fig:FermiSeas}.

The neutron and proton Fermi layers can be calculated from the
difference of two Fermi spheres with different radii defined as,
\begin{equation}
  n_n = \frac{1}{3\pi^2} \left( k_{F_n}^3 - k_{F_{n}^{\min}}^3 \right) \, , \,\,
  n_p = \frac{1}{3\pi^2} \left( k_{F_p}^3 - k_{F_{p}^{\min}}^3 \right) \, , \label{eq:nnp}
\end{equation}
where $k_{F_n^{\min}}$ and $k_{F_p^{\min}}$ are the lower bound of the
nucleon shell, see Fig.~\ref{fig:FermiSeas}.
Injecting Eq.~(\ref{eq:nnuc}) into $n_i=x_i n_N$ ($i=n$, $p$) and
identifying with Eqs.~\eqref{eq:nnp}, we obtain
\begin{equation}
  k_{F_n}^3 = (1+\delta_N) k_{F_N}^3\, , \hspace{0.5cm} k_{F_p}^3 = (1-\delta_N) k_{F_N}^3\, ,
  \label{eq:kfnp}
\end{equation}
as well as
\begin{eqnarray}
  k_{F_n^{\min}}^3 &=& (1+\delta_N) \left( N_c k_{F_Q}\right)^3\, ,  \\
  k_{F_p^{\min}}^3 &=& (1-\delta_N) \left( N_c k_{F_Q}\right)^3\, .
\end{eqnarray}
Note that $k_{F_n^{\min}}$ and $k_{F_p^{\min}}$ can be re-expressed as
$k_{F_i}^{\min}=(2 x_i)^{1/3} N_c k_{F_Q}=N_c(3\pi^2 x_i n_Q)^{1/3}$, for $i=n$, $p$.

At low densities, in the absence of quarks $k_{F_Q}=0$, so neutrons and
protons occupy entirely their Fermi spheres with radii given by
Eqs.~\eqref{eq:kfnp}. It is interesting to note that
Eqs.~\eqref{eq:kfnp} are identical in the presence or the absence of quarks.

Now that the Fermi spheres and the Fermi shells for $n$, $p$, $u$ and $d$
particles are well defined, all thermodynamical quantities can be
determined, e.g. the energy of the ground state, the pressure, the
chemical potentials, or the sound speed.

\subsection{Energy density and derivatives}

\begin{table*}[t]
  \centering
  \setlength{\tabcolsep}{6pt}
  \renewcommand{\arraystretch}{1.5}
  \begin{tabular}{cccccccccccccc}
    \hline
    \hline
    Model & $E_\sat$ & $E_\sym$ & $n_\sat$ & $L_\sym$ & $K_\sat$ & $K_\sym$ & $Q_\sat$ & $Q_\sym$ & $Z_\sat$ & $Z_\sym$ & $m^*/m$ & $\Delta m^*/m$ & $b_\sat$ \\
          & MeV & MeV & $\fmiq$ & MeV & MeV & MeV & MeV & MeV & MeV & MeV \\
    \hline
    SLy4$_\mm^\nr$ & -15.97 & 32.01 & 0.1595 & 46 & 230 & -120 & -225 & 400 & -443 & -690  & 0.69 & -0.19 & 6.90 \\
    SLy4$_\mm^\rl$ & -15.97 & 32.01 & 0.1595 & 46 & 230 & -120 & -225 & 400 & -443 & -690  & 1.0 & 0.0 & 6.90 \\
    \hline
    \hline
  \end{tabular}
  \caption{Parameters of the SLy4 meta-model used in the
    non-relativistic (NR) case for the description of nuclear matter
    and in the relativistic case (RL), where only relativistic kinematic is considered, for the quarkyonic matter.}
  \label{tab:mm}
\end{table*}

The energy density of quarkyonic matter is given by
\begin{eqnarray}
  \rho_B &=& \rho_N + \rho_Q \, ,
\end{eqnarray}
where the nucleon and quark terms are given by:
\begin{eqnarray}
  \rho_N &=& 2\!\!\sum_{i=n,p}\!\int_{k_{F_i}^{\min}}^{k_{F_i}} \!\! \frac{d^3k}{(2\pi)^3}\sqrt{k^2+M_N^2} + V_N(n_n,n_p) \, , \\
  \rho_Q &=&2\!\!\sum_{q=u,d}\!\!N_c\!\!\int_{0}^{k_{F_q}} \!\!\! \frac{d^3k}{(2\pi)^3}\sqrt{k^2+M_Q^2} \, ,
\end{eqnarray}
with the nuclear residual interaction given by the meta-model~\cite{Margueron:2018a},
\begin{eqnarray}
  V_N(n_n,n_p) = \sum_\alpha  \frac{1}{\alpha!} c_\alpha(\delta_N) x^\alpha u_\alpha(x) \, ,
\end{eqnarray}
where $x=(n_N-n_{\sat})/(3n_{\sat})$,
$c_\alpha(\delta_N)=c_{\sat,\alpha} + c_{\sym,\alpha} \delta_N^2$ and
$u_\alpha(x)=1-(-3x)^{N+1-\alpha}\exp(-b \, n_N/n_{\sat})$.
The coefficients $c_{\sat,\alpha}$ and $c_{\sym,\alpha}$ are related to
the empirical parameters, e.g. $E_{\sat}\approx -16$~MeV,
$n_{\sat}\approx 0.16$~fm$^{-3}$ and $K_{\sat}\approx 230$~MeV in
nuclear matter, considering the relativistic extension of the
meta-model (for the kinetic term only)~\cite{RMM}. The present calculation is based on the nucleon
Skyrme interaction SLy4, on which the meta-model is adjusted. We
consider the parameters given in table~\ref{tab:mm}.

The binding energy density $\epsilon_B$ is defined as
\begin{eqnarray}
  \epsilon_B (k_{F_N},\delta_N) = \rho_B (k_{F_N},\delta_N) - M_N n_B \, ,
\end{eqnarray}
and the binding energy per baryon number is
\begin{eqnarray}
  e_B (k_{F_N},\delta_N) = \epsilon_B(k_{F_N},\delta_N) / n_B \, .
\end{eqnarray}

The other quantities as the chemical potentials, pressure and sound
velocities are computed using the usual definitions, see
Ref.~\cite{Margueron:2018a,Margueron:2018b} for more details.

\subsection{Results}
\label{sec:res}

\begin{figure*}[t]
  \begin{center}
    \includegraphics[width=1.6\columnwidth]{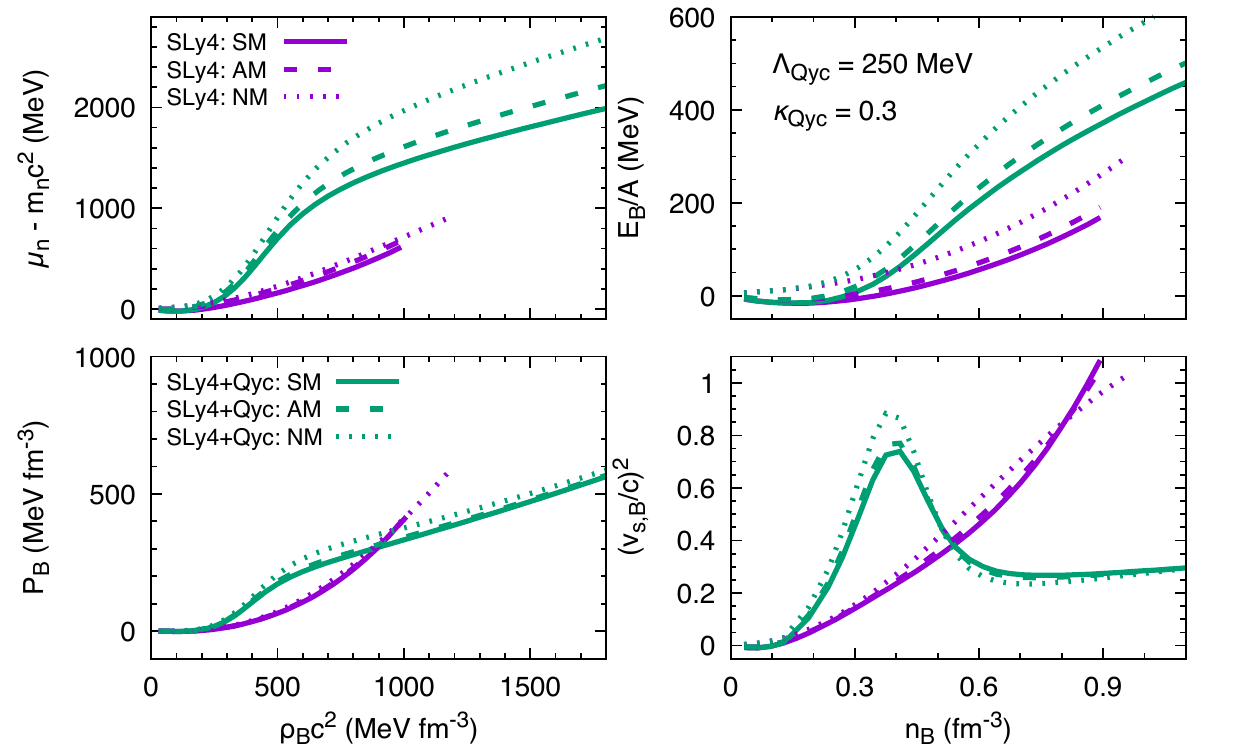}
    \caption{Baryon chemical potential $\mu_B$, energy per particle
      $E_B/A$, pressure $P_B$ and sound speed $(v_{s,B}/c)^2$ in
      symmetric (SM, $\delta_N=0.0$), asymmetric (AM, $\delta_N=0.5$), and neutron matter 
      (NM, $\delta_N=1.0$) for
      $\Lambda_{\Qyc}=250$~MeV and $\kappa_{\Qyc}=0.3$. }
    \label{fig:am:qic-250}
  \end{center}
\end{figure*}

In this section we compare a pure nucleon model for matter properties
against quarkyonic models constructed on top of the same nucleon model.
The choice of SLy4 is influenced by the fact that this pure nucleon
model reproduces most of the recent observational data, such as the
tidal deformability from GW170817 or the NICER x-ray observation of
PSR J0030+0451. We then explore the influence of the quarkyonic model
parameters $\Lambda_\Qyc$ and $\kappa_\Qyc$.

The neutron chemical potential $\mu_n$, the energy per particle
$E_B/A$, the baryon pressure $P_B$ and baryon sound speed
$(v_{s,B}/c)^2$ are shown in Fig.~\ref{fig:am:qic-250} for SM (solid
lines, $\delta_N=0$), AM (dashed lines, $\delta_N=0.5$) and NM (dotted
lines, $\delta_N=1$).
The quarkyonic model parameters are fixed to be $\Lambda_{\Qyc}=250$~MeV 
and $\kappa_{\Qyc}=0.3$.
The predictions for the quarkyonic phase (green lines) are confronted
to the ones for the pure nucleon phase (magenta lines).
The model predictions are stopped when causality is violated. 
%
%
The sound velocity in quarkyonic matter has a peak at around
$n_B\approx 0.4$~fm$^{-3}$, as shown in Ref.~\cite{McLerran:2019} for
SM and NM, and confirmed here for AM.
The position of the peak is almost independent of the isospin
asymmetry, but the peak is a bit more pronounced in NM compared to SM.
For the chosen parameters, the sound speed predicted by the quarkyonic model
at high density reaches a value close to the conformal limit, i.e. $1/3$.
The bump in the sound speed density profile present in quarkyonic matter
impacts the pressure, the chemical potential and the binding energy.
These thermodynamical quantities are strongly increased for densities 
where the sound speed is maximal, and they are softer at higher densities.
The softening is such that the pressure of quarkyonic matter crosses the 
pure nucleon one at high density, see Fig.~\ref{fig:am:qic-250}.
The softening of the EoS is also predicted by usual construction of 
first order phase transitions from nucleon to quark matter, and the 
interesting feature of the quarkyonic model is the stiffening of the EoS
at low densities, where it really matters for NS, before the softening
at high density. 
The region of importance for NS properties coincides mostly with the 
densities where the pressure is stiff.
This is the reason why this model is of particular interest for the 
phenomenology of compact stars.

The increase of the chemical potential in the cross-over region can also
be explained from the behavior of the nucleon Fermi momentum, which
can be traced down from Eq.~\eqref{eq:nnuc} and re-written as,
\begin{equation}
k_{F_N}^3 =  \frac{3\pi^2}{2}  \Big[  n_N + N_c^3 n_Q \Big]  \, ,
\end{equation}
showing that the quark contribution to the nucleon Fermi momentum is
strongly enhanced by the factor $N_c^3$. 
As the density increases however, the quark component of matter
becomes more and more dominant and the softening actually occurs.

The conclusion of this first set of results is that the generic
features of the quarkyonic model predicted by McLerran and
Reddy~\cite{McLerran:2019} are preserved in our extension of the
quarkyonic model for AM, and we can predict similar features in NM
with the same parameters as the one in SM.
Our results are also in qualitative agreement with the ones found by
Zhao and Lattimer~\cite{Zhao:2020} as well as Jeong McLerran and
Sen~\cite{Jeong:2020} where different nuclear potential were used.

\section{Quarkyonic model at $\beta$-equilibrium}
\label{sec:qbm}

We are now constructing the beta-equilibrium solutions, which describe the ground
state of dense matter existing in the core of compact stars.
In cold catalyzed NS, matter components are determined from the
$\beta$-equilibrium equations,
\begin{eqnarray}
  \mu_n-\mu_p &=& \mu_e \, , \\
  \mu_e &=& \mu_\mu \, ,
\end{eqnarray}
and charge neutrality,
\begin{eqnarray}
  n_e+n_\mu+\frac 1 3 n_d = n_p + \frac 2 3 n_u \, .
  \label{eq:cn}
\end{eqnarray}
At fixed total density, these three equations allows the determination
of three variables: the isospin asymmetry $\delta_N$ and the electron
and muon densities, $n_e$ and $n_\mu$.

Note that the charge neutrality condition~\eqref{eq:cn} in NM becomes
$n_d=2 n_u$, which coincides with the relation between $u$ and $d$
quark Fermi momenta, $k_{Fd}=2^{1/3}k_{Fu}$, imposed in NM in
\cite{McLerran:2019}.

\begin{figure}[tb]
  \begin{center}
    \includegraphics[width=0.99\columnwidth]{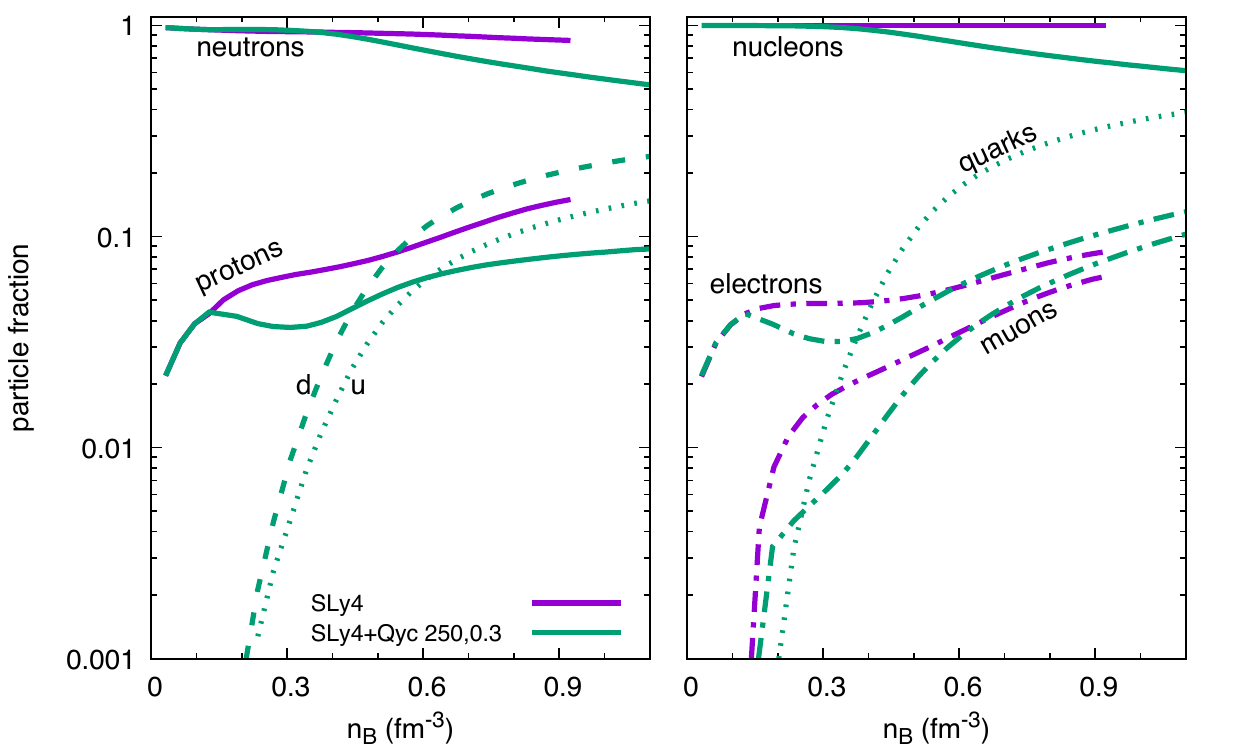}
    \caption{Particle fractions for the SLy4 nucleon model and the
      quarkyonic model taking $\Lambda_{\Qyc}=250$~MeV and
      $\kappa_{\Qyc}=0.3$.}
    \label{fig:beta:xi}
  \end{center}
\end{figure}

The particle fractions in dense matter are shown in
Fig.~\ref{fig:beta:xi} for the SLy4 nucleon model (magenta) and the
quarkyonic model taking $\Lambda_{\Qyc}=250$~MeV and
$\kappa_{\Qyc}=0.3$ (green).
On the left panel of Fig.~\ref{fig:beta:xi} are shown only the baryon
contributions, $n$, $p$, $d$ and $u$, while on the right panel, the
contributions of the nucleons, quarks, electrons and muons are
represented.
In the crossover region, neutron and proton densities are reduced
compare to the original nucleon model while the amount of quarks
increases.
In particular, we observe that the fraction of protons is reduced in
the quarkyonic model such that it remains below the dURCA
threshold ($\gtrsim 1/9$\% in the presence of muons~\cite{Lattimer:1991} ).

\begin{figure}[tb]
  \begin{center}
    \includegraphics[width=0.99\columnwidth]{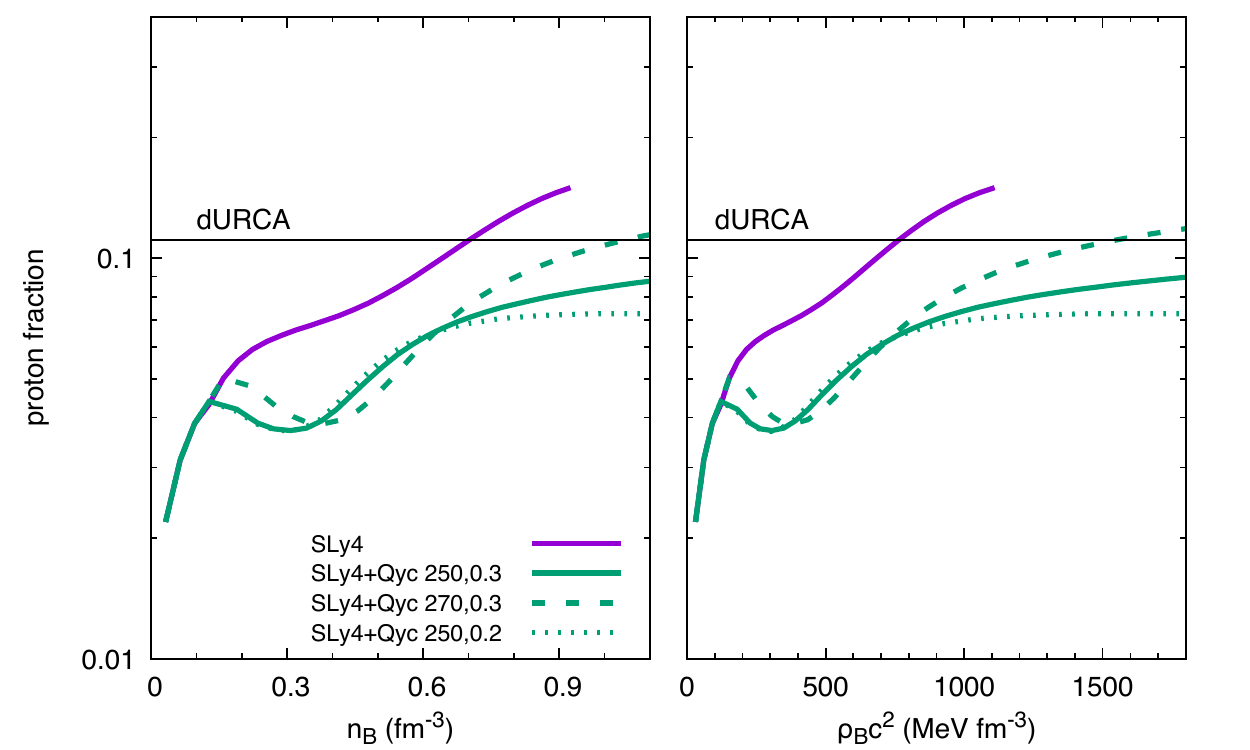}
    \caption{Comparison of the proton fraction in the pure nucleon
      model with the prediction of the quarkyonic model for various
      sets of the parameters ($\Lambda_{\Qyc}$, $\kappa_{\Qyc}$):
      ($250,0.3$), ($270,0.3$) and ($250,0.2$). Left: function of the
      particle density in $\fmiq$, right: function of the energy
      density $\rho_B c^2$ in MeV~$\fmiq$.}
    \label{fig:beta:yp}
  \end{center}
\end{figure}

\begin{figure*}[tb]
  \begin{center}
    \includegraphics[width=1.6\columnwidth]{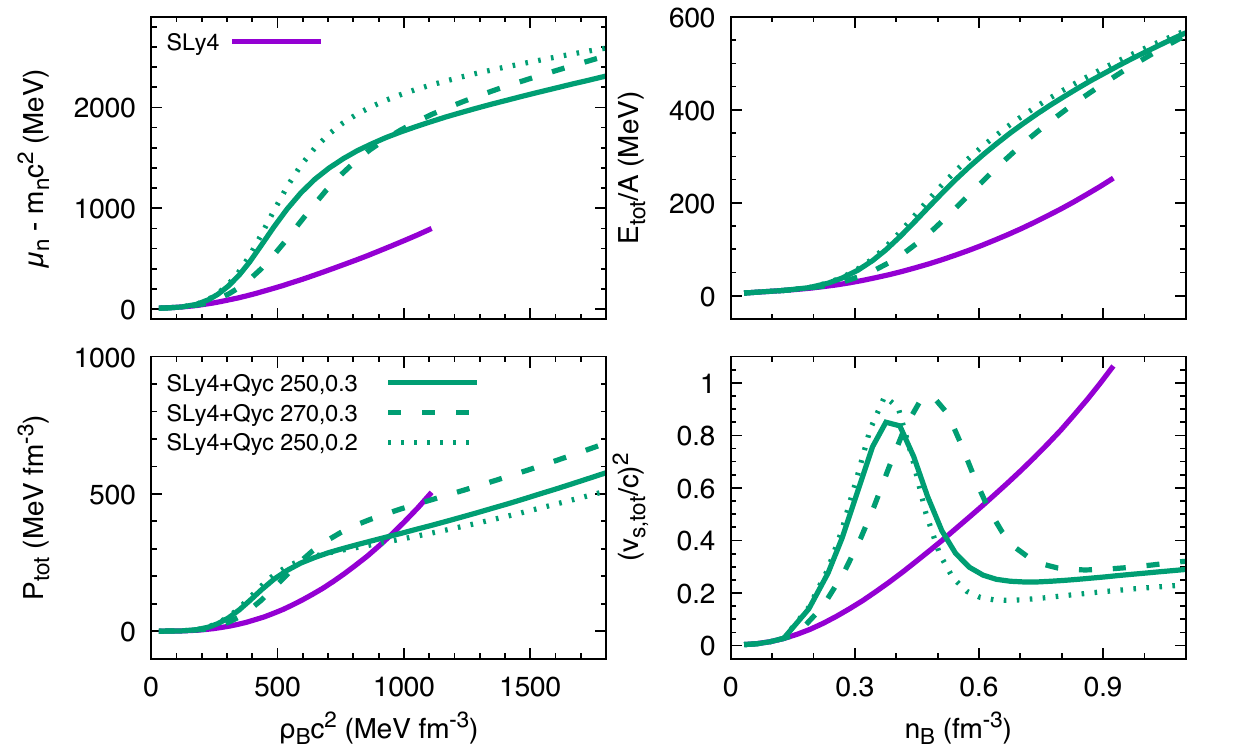}
    \caption{Neutron chemical potential $\mu_n$, total energy per
      particle $E_{tot}/A$, total pressure $P_{tot}$ and total sound
      speed $(v_{s,tot}/c)^2$ at $\beta$-equilibrium for the
      parameters ($\Lambda_{\Qyc}$, $\kappa_{\Qyc}$), 
      same choice as in Fig.~\ref{fig:beta:yp}.}
    \label{fig:beta:qic}
  \end{center}
\end{figure*}

In order to investigate the role of the parameters of the quarkyonic
model on the proton fraction, we show in Fig.~\ref{fig:beta:yp} a
comparison for different choices for the parameters ($\Lambda_{\Qyc}$,
$\kappa_{\Qyc}$): ($250,0.3$), ($270,0.3$) and ($250,0.2$).
Increasing $\Lambda_{\Qyc}$ from 250~MeV to 270~MeV induces an
increase of the proton fraction at high density, which passes the
dURCA threshold, but at higher density. In terms of energy density $\rho_B c^2$,
right panel in Fig.~\ref{fig:beta:yp}, the dURCA threshold is pushed 
even higher, since the quarkyonic model predicts larger energy densities 
than the pure nucleon one.

Finally, the thermodynamical properties of quarkyonic matter at
$\beta$-equilibrium are shown in Fig.~\ref{fig:beta:qic}: the neutron
chemical potential $\mu_n$, the total energy per particle $E_{tot}/A$,
the equation of state $P(\rho_B c^2)$ and the total sound speed.
The effect of varying the parameters of the quarkyonic matter is also
shown. The largest impact is observed when the parameter
$\Lambda_{\Qyc}$ in increased from 250~MeV to 270~MeV. Increasing
$\Lambda_{\Qyc}$ has the effect of pushing the onset of the first
quarks at higher density, as it can also be observed in
Fig.~\ref{fig:beta:qic}. As a consequence, increasing $\Lambda_{\Qyc}$
makes quarkyonic matter more and more repulsive, except at low density
where the larger $\Lambda_{\Qyc}$ the softer the EoS, as discussed
previously.
The effect of changing $\kappa_{\Qyc}$ is smaller. It was tuned in the
original paper by McLerran and Reddy~\cite{McLerran:2019} to the
conformal limit for the sound speed.

\section{Quarkyonic stars}
\label{sec:ns}

The structure of non-rotating neutron stars is provided by the
solution of the spherical hydrostatic equations in general relativity,
also named the Tolmann-Oppenheimer-Volkof equations~\cite{TOV},
\begin{eqnarray}
  \frac{dm(r)}{dr} &=& 4\pi r^2\rho(r),   \label{eq:tov} \\
  \nonumber  \frac{dP(r)}{dr} &=& -\rho(r) c^2\Bigg(1+\frac{P(r)}{\rho(r) c^2} \Bigg)\frac{d\Phi(r)}{dr}, \\
  \nonumber  \frac{d\Phi(r)}{dr} &=& \frac{Gm(r)}{c^2 r^2}\Bigg(1+\frac{4\pi P(r) r^3}{m(r) c^2} \Bigg)\Bigg(1-\frac{2Gm(r)}{r c^2} \Bigg)^{-1},
\end{eqnarray}
where $G$ is the gravitational constant, $c$ the speed of light,
$P(r)$ the total pressure, $m(r)$ the enclosed mass, $\rho(r)=\rho_B(r)$ is the
total mass-energy density and $\Phi(r)$ the gravitational field.
$P$ and $\rho$ have contributions from both the baryons ($P_B$,
$\rho_B$) and the leptons ($P_L$, $\rho_L$).

The four variables ($m$, $\rho$, $P$, $\Phi$) are obtained from the
solution of the three TOV equations~(\ref{eq:tov}) and the EoS for the
quarkyonic matter. In the present calculation, a crust EoS is smoothly
connected to the core EoS as described in Ref.~\cite{Margueron:2018b}.
The tidal deformability $\Lambda_{GW}$ induced by an external quadrupole field is
expressed in terms of the Love number $k_2$ as
$\Lambda_{GW}=2k_2/(3C^5)$, where the compactness is $C=GM/(Rc^2)$, and
$k_2$ is calculated from the pulsation equation at the surface of
NS~\cite{Hinderer:2008,Flanagan:2008},
\begin{eqnarray}
k_2 &=& \frac{8 C^5}{5}\left( 1-2C\right)^2\left(2-y_R+2C(y_R-1)\right)\nonumber \\
&&\hspace{-1cm}\times\Big(2C(6-3y_R+3C(5y_R-8))\nonumber \\
&&\hspace{-1cm}+4C^3\left(13-11y_R+C(3y_R-2)+2C^2(1+y_R)\right)\nonumber \\
&&\hspace{-1cm}+3(1-2C)^2\left(2-y_R+2C(y_R-1)\right)\ln(1-2C) \Big)^{-1} \!\!\!\! ,
\end{eqnarray}
where $y_R$ is the value of the $y$ function at radius $R$, $y_R=y(r=R)$, and $y(r)$ is the solution of
the following differential equation,
\begin{equation}
r\frac{dy}{dr}+y^2+yF(r)+r^2Q(r)=0\, ,
\end{equation}
with the boundary condition $y(0)=2$ and the functions $F(r)$ and $Q(r)$ defined as,
\begin{eqnarray}
F(r)&=&\frac{1-4\pi r^2 G[\rho(r)-P(r)]/c^4}{1-2M(r)G/(rc^2) } \, ,\\
r^2Q(r) &=& \frac{4\pi r^2 G}{c^4}\left(5\rho(r)+9P(r)+\frac{\partial \rho(r)}{\partial P(r)}[\rho(r)+P(r)]\right)\nonumber \\
&&\hspace{-1.2cm}\times\left(1-2M(r)G/(rc^2)\right)^{-1} - 6\left( 1-2M(r)G/(rc^2) \right)^{-1} \nonumber \\
&&\hspace{-1.2cm}-\frac{4 G^2}{r^2c^8}\left( M(r)c^2+4\pi r^3P(r)\right)^2\left(1-2M(r)G/(rc^2)\right)^{-2} \, .\nonumber \\
\end{eqnarray}

The NS moment of inertia is obtained from the slow rotation approximation~\cite{Hartle:1967,Morrison:2004} as
\begin{equation}
I = \frac{8\pi}{3}\int_0^R dr r^4 \rho(r)\left(1+\frac{P}{\rho(c)c^2}\right)\frac{\bar{\omega}}{\Omega}e^{\lambda-\Phi} \, ,
\end{equation}
where $\bar{\omega}$ is the local spin frequency, which represents the general relativistic correction to the asymptotic angular momentum $\Omega$ and $\lambda$ is defined as $\exp(-2\lambda)=1-Gm/(rc^2)$.

As usual, for a given EoS the family of solutions is parameterized by
the central density or pressure or enthalpy.
The EoS are characterized by their evolution in the mass-radius
diagram, where both masses and radii of compact stars could in
principle be measured, as discussed in our introduction, see also
Ref.~\cite{Watts:2015}.

\begin{figure*}[tb]
  \begin{center}
    \includegraphics[width=1.8\columnwidth]{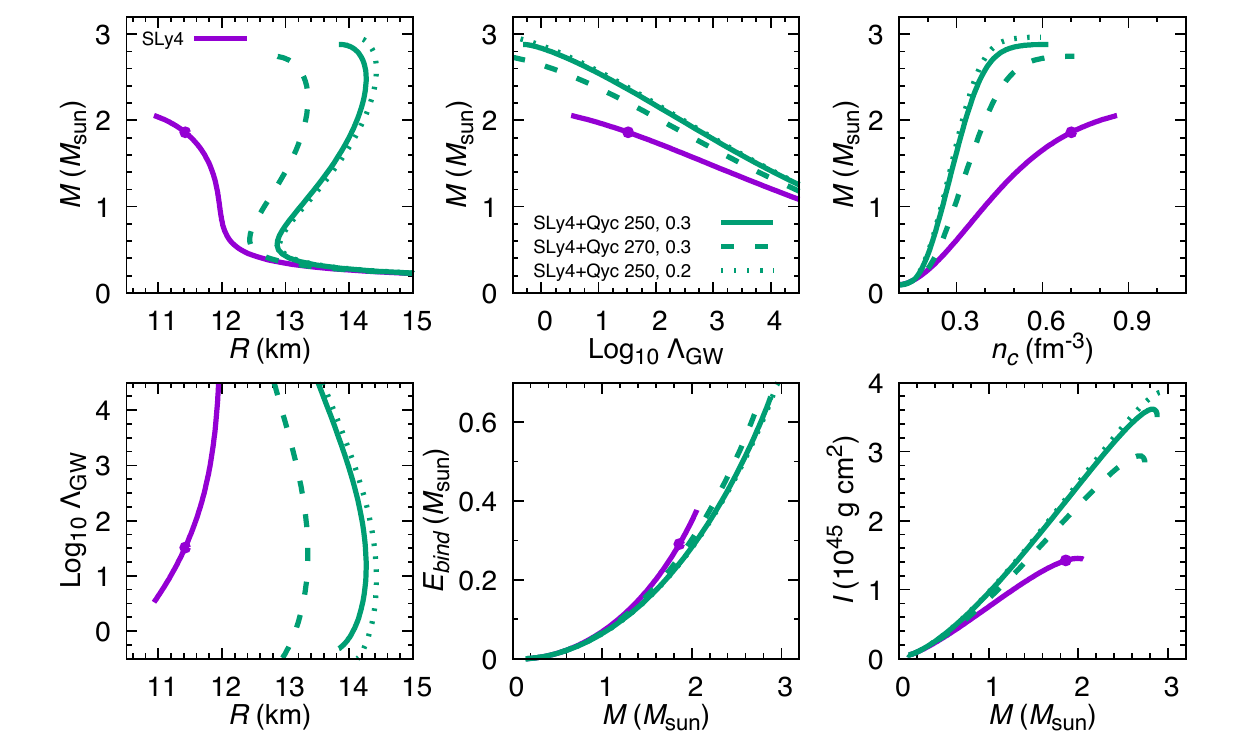}
    \caption{Neutron star properties (mass $M$, radius $R$, tidal
      deformability $\Lambda_{GW}$, central density $n_c$, binding energy $E_{bind}=M_b-M$, 
      where $M_b$ is the baryon mass, and the moment of inertia $I$) for
      quarkyonic matter with $\Lambda_{\Qyc}$ and $\kappa_{\Qyc}$ fixed as in Fig.~\ref{fig:beta:yp}
      (green curves) 
      and nucleon matter (solid magenta curve). }
    \label{fig:ns}
  \end{center}
\end{figure*}

We show in Fig.~\ref{fig:ns} the predictions for the mass $M$, 
radius $R$, tidal deformability $\Lambda_{GW}$, central density $n_c$,
binding energy $E_{bind}$ and moment of inertia $I$ associated to 
various quarkyonic EoS with ($\Lambda_{\Qyc}$, $\kappa_{\Qyc}$): 
($250,0.3$), ($270,0.3$) and ($250,0.2$) (green lines). 
These predictions are confronted to the ones for a nucleon EoS
(solid magenta line).

The impact of quarkyonic matter on the mass-radius relation is huge,
as already noticed in Refs.~\cite{McLerran:2019,Jeong:2020}. While the
maximum mass for SLy4 is reached at 2.03~M$_\odot$, the quarkyonic
stars almost reach 3~M$_\odot$. These is also a large impact on the
radius: the 1.4~M$_\odot$ radius $R_{1.4}$ of the pure nucleon model
is about 11.5~km, while it is pushed up to 13-14~km in the quarkyonic
model. Quarkyonic stars can therefore be much more massive than pure
nucleon ones, and are also bigger in size.
Quarkyonic stars have also different tidal deformabilities
$\Lambda_{GW}$ compared to the pure nucleon case. 
For the same mass, the quarkyonic stars have larger $\Lambda_{GW}$, 
and at fixed $\Lambda_{GW}$, they have larger radii. At a fixed central
density (in $\fmiq$), quarkyonic stars are much more massive than the
pure nucleon model we considered. This is an effect of the repulsion
observed for the pressure in Fig.~\ref{fig:beta:qic}.
For the same mass, quarkyonic stars have a slightly lower $E_{bind}$ 
compared to the associated nucleonic star, they however have a larger 
moment of inertia.

One can estimate the influence of the parameters $\Lambda_{\Qyc}$ and 
$\kappa_{\Qyc}$. 
As observed in previous figures, the parameter $\kappa_{\Qyc}$ is almost 
not influential at all, while $\Lambda_{\Qyc}$ is much more critical. 
By increasing $\Lambda_{\Qyc}$, the realization of quarkyonic matter is pushed up
and the EoS get closer to the pure nucleon case. The quarkyonic star
therefore also get a bit closer to the neutron star as
$\Lambda_{\Qyc}$ increases. The journey in the mass-radius diagram is
therefore very much controlled by the parameter $\Lambda_{\Qyc}$. By
decreasing $\Lambda_{\Qyc}$ the quarkyonic star gets bigger and bigger
compared to the NS, as well as it gets more and more massive. 
In the future, the observation of several points in mass and radius, e.g. from
NICER observations, will thus be very useful for the determination of 
the parameter $\Lambda_{\Qyc}$.

Finally we discuss the dURCA threshold, represented in the curves by
the solid circle. Only the pure nucleon case reaches the dURCA
threshold, and it happens close to 2~M$_\odot$. Even the quarkyonic model with
($\Lambda_{\Qyc}$, $\kappa_{\Qyc}$)=($270,0.3$), which satisfies the
dURCA condition at high density, see Fig.~\ref{fig:beta:yp}, reaches
the unstable branch before it gets to the dURCA density. It is thus
more difficult to reach the dURCA condition with quarkyonic stars.
The same conclusion was also obtained by Zhao and Lattimer~\cite{Zhao:2020}
with their version of quarkyonic matter.

\begin{figure}[tb]
  \begin{center}
    \includegraphics[width=0.99\columnwidth]{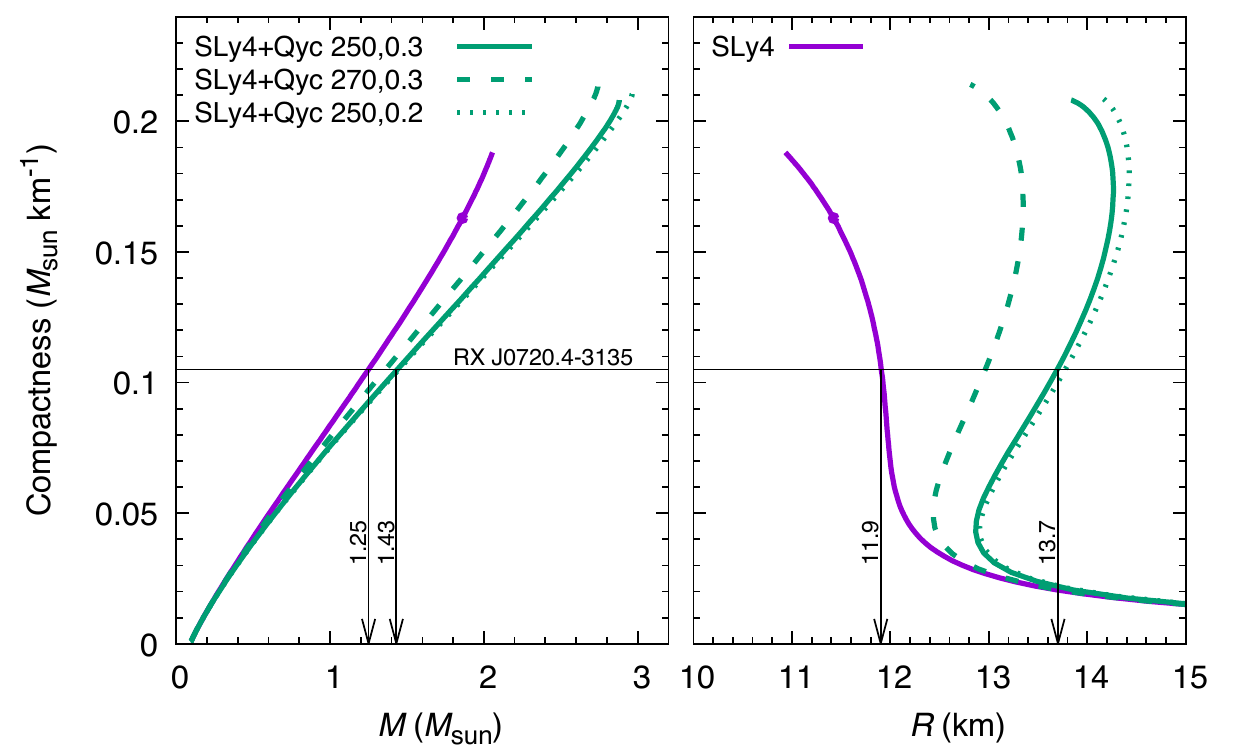}
    \caption{Compactness $(M/M_\odot)/(R/\hbox{km})$ as function of the
    mass $M/M_\odot$ (left) and radius (right) for various
      sets of the parameters $\Lambda_{\Qyc}$ and $\kappa_{\Qyc}$ fixed to be 
      as in Fig.~\ref{fig:beta:yp}.}
    \label{fig:tov:c}
  \end{center}
\end{figure}

We show in Fig.~\ref{fig:tov:c} the NS compactness defined as $(M/M_\odot)/(R/\hbox{km})$, where $R$ is expressed in km, function of the mass (left panel) and of the radius (right panel). 
The compactness of the isolated NS RX J0720.4-3135 has been extracted from observations and estimated to be 0.105$\pm$0.002~\cite{Hambaryan:2017}.
Reporting this value in the left panel of Fig.~\ref{fig:tov:c} we observe that the nucleonic EoS SLy4 suggests that the mass of the pulsar is 1.25~$M_\odot$, compatible with observed masses but close to their lower limit~\cite{Ozel:2016b}, while the quarkyonic model suggest higher masses, up to 1.43~$M_\odot$, which is compatible with the canonical NS mass.
Note that a Bayesian exploration of nucleonic models has predicted a centroid of about $1.33~M_\odot$~\cite{Margueron:2018b}. 
This value is slightly larger than the one suggested by SLy4 EoS, but is still lower that the canonical mass.
On the right panel of Fig.~\ref{fig:tov:c} the radii associated to the observed compactness are also reported. While the SLy4 EoS favors 11.9~km, the quarkyonic stars point towards larger radii, up to 13.6~km in the upper case.
So we conclude that for a fixed value of the compactness, Fig.~\ref{fig:tov:c} shows that quarkyonic stars have larger masses  and radii than NS.

\begin{figure}[tb]
  \begin{center}
    \includegraphics[width=0.99\columnwidth]{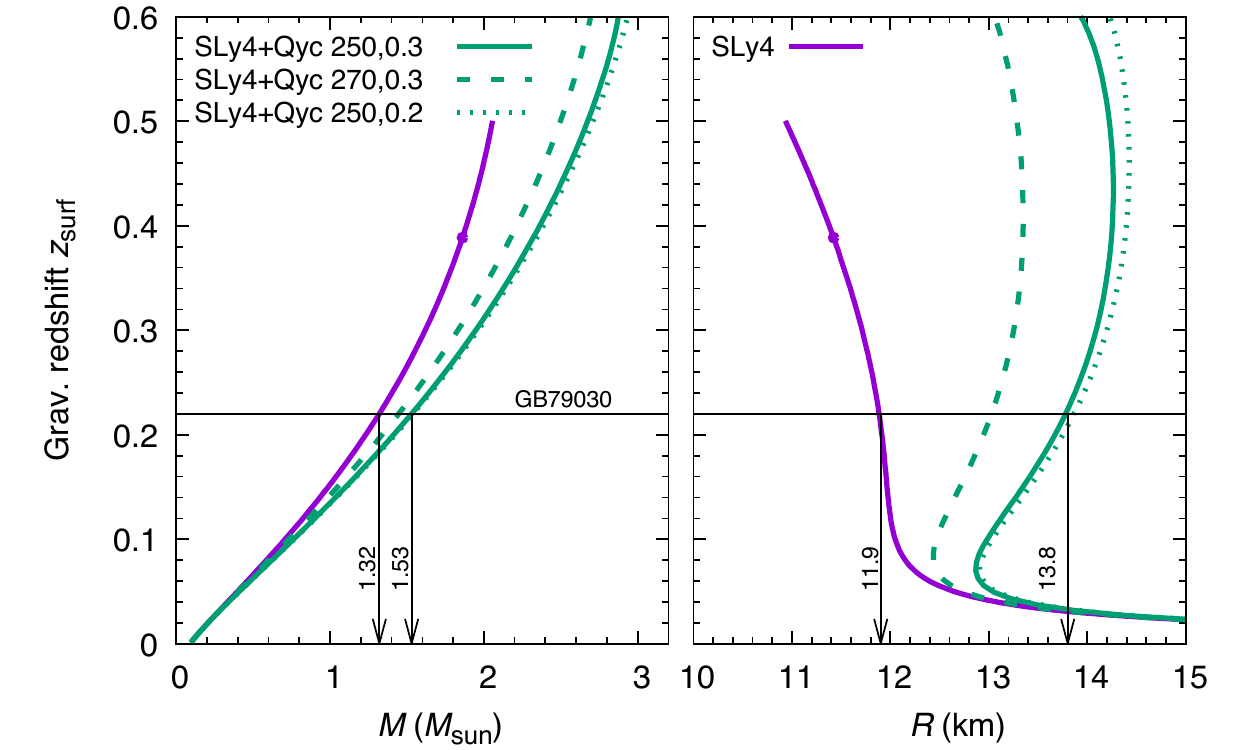}
    \caption{Gravitational redshift $z_\mathrm{surf}$ as function of the
    mass $M/M_\odot$ (left) and radius (right) for various
      sets of the parameters $\Lambda_{\Qyc}$ and $\kappa_{\Qyc}$ fixed to be 
      as in Fig.~\ref{fig:beta:yp}.}
    \label{fig:tov:z}
  \end{center}
\end{figure}

The gravitational redshift $z_\mathrm{surf}$ associated to the radial emission of photons from the surface, which is detected by a distant observer is defined as $z_\mathrm{surf}=\left(1-2GM/(Rc^2)\right)^{-1/2}-1$.
In Fig.~\ref{fig:tov:z} we show $z_\mathrm{surf}$ versus the stellar mass (left panel) and versus the stellar radius (right panel).
The emission line feature of the gamma ray burst GB790305, assumed to originate from the $e^- e^+$ annihilation (observed peak at 430 keV, line width 150$\pm$20 keV), and assuming thermal nature of line broadening and taking due account of the thermal blueshift, leads to the observational constraint $z_\mathrm{surf}^\mathrm{GB790305}=0.22$~\cite{Mazets:1982,Higdon:1990}, which is reported in Fig.~\ref{fig:tov:z}.
We also deduce from this observational data that SLy4 would favor typical masses of the order of 1.32~$M_{\odot}$, low masses, while the quarkyonic star built on the same nucleonic star would point towards 1.53~$M_{\odot}$, closer to the canonical NS mass.
The radius of extracted from SLy4 would be 11.9~km, while quarkyonic star would point towards larger radii, up to about 13.8~km.

\begin{figure}[tb]
  \begin{center}
    \includegraphics[width=0.99\columnwidth]{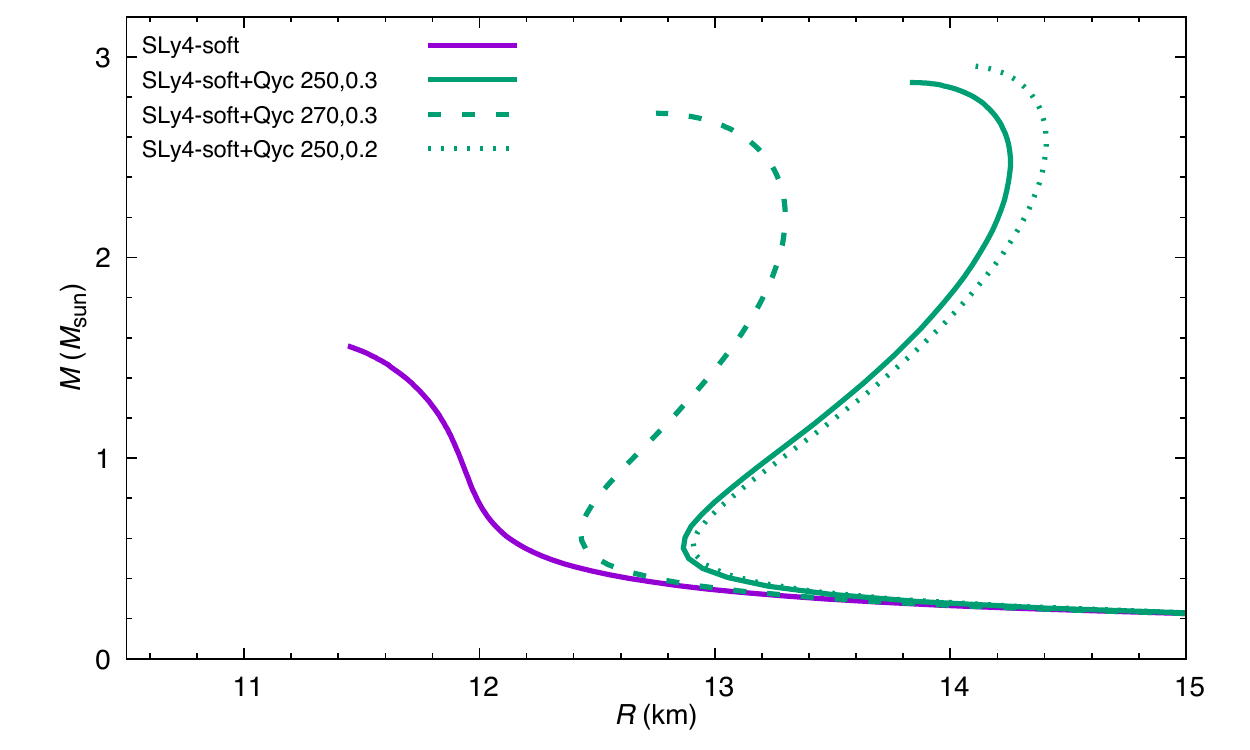}
    \caption{Mass-radius relation for SLy4-soft nuclear interaction and for the quarkyonic model with various
      sets of the parameters $\Lambda_{\Qyc}$ and $\kappa_{\Qyc}$ fixed to be 
      as in Fig.~\ref{fig:beta:yp}.}
    \label{fig:tov:soft}
  \end{center}
\end{figure}

Finally, we construct the quarkyonic model on top of a nucleonic model which does not reach the observation constraint of about $2M_\odot$. To do so, we reduce the value of $Z_{\sym}$ from the SLy4 nucleonic model by 300~MeV, see table~\ref{tab:mm}. The nucleonic model is shown in Fig.~\ref{fig:tov:soft} under the label SLy4-soft (solid magenta line) while the quarkyonic models are shown for the same three cases as before.
With this example, we show that the crossover transition to quark matter, as described by the quarkyonic approach, can bring enough repulsion to reach large  maximum mass, even if the model for the nucleonic part cannot satisfy the observed requirement that the maximal mass of NS should be above about $2M_\odot$.

\section{Conclusions}
\label{sec:conclusions}

We have proposed an extension of the original
quarkyonic model from Ref.~\cite{McLerran:2019} to AM, where the
original quarkyonic model for SM is recovered as a limit.
Our extension assumes (i) that the description of the quark Fermi see and
nucleon shell is globally isoscalar and (ii) that the isospin-flavor asymmetry 
in the quark and nucleon phases is fixed. These assumptions root into the 
concept of the quarkyonic model where nucleons result from the strong
confining force, which strength is large close to the Fermi level.
The assumption (i) allows us to smoothly connect to the quarkyonic model in 
SM, and suggests a description of NM quite comparable -- at least qualitatively -- 
to the original one suggested by McLerran and Reddy~\cite{McLerran:2019}.
By fixing the isospin/flavor asymmetry in the nucleon and quark phases,
assumption (ii), the properties of isospin asymmetric quarkyonic matter can be 
entirely determined from the nucleon Fermi momentum $k_{F_N}$ and the isospin
asymmetry $\delta_N$.

NS matter at $\beta$-equilibrium is then been calculated and it provides
qualitatively similar results as in the original model~\cite{McLerran:2019}.
It is also in agreement with other extensions in asymmetric 
matter~\cite{Zhao:2020, Jeong:2020}, while being based on different
assumptions.
In our model, quarkyonic stars are larger and heavier than the associated
NS, and the parameter $\Lambda_{\Qyc}$ is playing a dominant role in
changing the radius and the mass of quarkyonic stars. This result is valid even 
if the nucleonic component is soft, e.g. too soft to reach $2M_\odot$.
The proton fraction at $\beta$-equilibrium is found to be reduced in the
quarkyonic matter, compared to the related pure nucleonic phase, which
potentially quench fast cooling -- based on dURCA process -- in massive compact stars.
The confrontation to a set of masses and radii, potentially obtained in future
observations like NICER or gravitational wave detections of in-spiral binary NS,
will potentially constrain $\Lambda_{\Qyc}$, as well as cooling scenarios.

In the future, we aim at incorporating the quarkyonic model in
systematical comparisons to observational data in order to better understand
the properties of dense matter. Extension of the present model to
finite temperature is also on our map for the near future, as well as
improving the isospin/flavor asymmetry relation.
Also, adding chiral symmetry consideration in the quarkyonic model,
taking into account the two most striking features of QCD, will
certainly be an interesting extension to study.

\acknowledgments

We thank K. S. Jeong, L. McLerran, S. Reddy, and A. Schmitt for fruitful discussions, 
as well as R. Stiele, R. Somasundaram, and A. Pfaff for careful checks of our
formalism.
J.M., H.H. and G.C. are supported by the CNRS/IN2P3 NewMAC project,
and benefit from PHAROS COST Action MP16214. All authors are grateful
to the LABEX Lyon Institute of Origins (ANR-10-LABX-0066) of the
\textsl{Universit\'e de Lyon} for its financial support within the
program \textsl{Investissements d'Avenir} (ANR-11-IDEX-0007) of the
French government operated by the National Research Agency (ANR).

\end{document}